\documentclass[aps,prx,twocolumn,notitlepage,superscriptaddress]{revtex4-1}

\usepackage[colorlinks=true, citecolor=magenta, linkcolor=blue, urlcolor=blue]{hyperref}
\usepackage{graphicx}
\usepackage{amsmath}
\usepackage{amssymb}
\usepackage{amsfonts}
\usepackage{xcolor}
\usepackage{tabularx,booktabs}
\usepackage{multirow}
\usepackage{tikz}
\usepackage{bibunits}
\usepackage{comment}

\usetikzlibrary{shapes.geometric, arrows}
\usetikzlibrary{decorations.markings}
\usetikzlibrary{decorations.pathmorphing}
\usetikzlibrary{decorations.pathreplacing}
\usetikzlibrary{shapes.symbols}

\allowdisplaybreaks

\begin{document}
% \begin{bibunit}[naturemag]
\bibliographystyle{naturemag}

\title{Dynamic Interfacial Quantum Dipoles in Charge Transfer Heterostructures}

\author{Ziyu Liu}
\affiliation{Department of Physics, Columbia University, New York, NY, USA}

\author{Emil Vi\~nas Bostr\"om}
\affiliation{Max Planck Institute for the Structure and Dynamics of Matter, Hamburg, Germany}
\affiliation{Nano-Bio Spectroscopy Group, Departamento de F\'isica de Materiales, Universidad del Pa\'is Vasco, 20018 San Sebastian, Spain}

\author{Dihao Sun}
\affiliation{Department of Physics, Columbia University, New York, NY, USA}

\author{Jordan Pack}
\affiliation{Department of Physics, Columbia University, New York, NY, USA}

\author{Matthew Cothrine}
\affiliation{Department of Materials Science and Engineering, University of
Tennessee, Knoxville, TN, USA.}

\author{Kenji Watanabe}
\affiliation{Research Center for Electronic and Optical Materials, National
Institute for Materials Science, Tsukuba, Japan.}

\author{Takashi Taniguchi}
\affiliation{Research Center for Materials Nanoarchitectonics, National Institute for Materials Science, Tsukuba, Japan.}

\author{David G. Mandrus}
\affiliation{Department of Materials Science and Engineering, University of
Tennessee, Knoxville, TN, USA.}
\affiliation{Materials Science and Technology Division, Oak Ridge National
Laboratory, Oak Ridge, TN, USA.}

\author{Angel Rubio}
\affiliation{Max Planck Institute for the Structure and Dynamics of Matter, Hamburg, Germany}
\affiliation{Initiative for Computational Catalysis, The Flatiron Institute, New York, NY, USA}

\author{Cory R. Dean}
\affiliation{Department of Physics, Columbia University, New York, NY, USA}
% \date{\today}

\begin{abstract}
\end{abstract}

\maketitle

%%%%%%%%%%%%%%%%%%%%%%%%%%%%%%%%%%%%%%%%%%%%%%%%%%%%%%%
%%%%%%%%%%%%%%%%%%%%%%%%%%%%%%%%%%%%%%%%%%%%%%%%%%%%%%%
%%%%%%%%%%%%%%%%%%%%%%%%%%%%%%%%%%%%%%%%%%%%%%%%%%%%%%%

\textbf{Hysteretic gate responses of two-dimensional material heterostructures serve as sensitive probes of the underlying electronic states and hold significant promise for the development of novel nanoelectronic devices. Here we identify a new mechanism of hysteretic behavior in graphene/$h$BN/$\alpha$-$\mathrm{RuCl_3}$ charge transfer field effect devices. The hysteresis loop exhibits a sharp onset under low temperatures and evolves symmetrically relative to the charge transfer equilibrium. Unlike conventional flash memory devices, the charge transfer heterostructure features a transparent tunneling barrier and its hysteretic gate response is induced by the dynamic tuning of interfacial dipoles originating from quantum exchange interactions. The system acts effectively as a ferroelectric and gives rise to remarkable tunability of the hysteretic gate response under external electrical bias. Our work unveils a novel mechanism for engineering hysteretic behaviors via dynamic interfacial quantum dipoles.}\\

Gate tunability plays an essential role in both understanding and exploiting the electronic properties of graphene-based devices~\cite{novoselov2005two,zhang2005experimental,zhou2021half,zhou2022isospin,lu2024fractional,li2024tunable,choi2025superconductivity, park2021tunable, kennes2021moire}. A critical characteristic associated with field effect control is the presence of hysteresis. In graphene field effect devices, gate hysteresis is frequently attributed to unintentional interfacial charge traps, adsorbed impurities and mobile ions \cite{wang2010hysteresis,joshi2010intrinsic,liao2010hysteresis,ju2014photoinduced}. Non-volatile memory structures intentionally introduce hysteresis through inclusion of a remote charge trapping layer, with the charge transfer dynamics determined by a combination of the dielectric spacer and trapping layer characteristics~\cite{hong2011graphene,bertolazzi2013nonvolatile,sup2013controlled}. Lattice scale distortions in moire heterostructures can yield a type of sliding ferroelectricity that is hysteretic under applied fields~\cite{vizner2021interfacial,yasuda2021stacking,weston2022interfacial}. A recent anomaly has been reported across several graphene-based devices, referred to as the “gate doesn’t work” effect, in which an unusual type of hysteretic charging behavior is observed. This effect has been linked to a possible interplay between remote trapping layers, emergent ferroelectricity, and electron interactions~\cite{zheng2020unconventional,yan2023moire,zheng2023electronic,waters2025anomalous} but still lacks a consensus understanding, and is difficult to reliably reproduce~\cite{skinner2023curious}.\\

Here we identify a new mechanism of hysteretic response in a charge transfer device. The main element is a three layer heterostructure composed of graphene/$h$BN/$\alpha$-$\mathrm{RuCl_3}$ (denoted as the GBR heterostructure), fabricated with a top and bottom gate. When gating from the $\alpha$-$\mathrm{RuCl_3}$ side (bottom gate), the heterostructure operates like a floating gate memory device with the $\alpha$-$\mathrm{RuCl_3}$ acting as the charge trapping layer. However, the associated hysteresis exhibits several unique characteristics: 1) at high temperature the bottom gate is fully screened, showing no ability to modulate the graphene density. Only below a threshold temperature, the bottom gate effect turns on and exhibits a hysteretic response whose character saturates with further cooling. 2) The hysteresis manifests under strong potential difference across the heterostructure due to the charge transfer between the graphene and $\alpha$-$\mathrm{RuCl_3}$ layers, driven by intrinsic work function difference. This severely bends the bands in the $h$BN spacer rendering it effectively transparent to tunneling. 3) The hysteresis behavior does not depend on the $h$BN spacer thickness, and indeed is observable even when removed entirely.\\

We ascribe the hysteresis to the emergence of local quantum dipoles that arise due to distortions of electron wavefunctions across the van der Waals interfaces, and form independent of charge transfer, structural transition or chemical reactions. The formation of an interfacial quantum dipole at the junction between two van der Waals materials has been identified previously both in experimental study~\cite{rizzo2023polaritonic} and on theoretical grounds~\cite{bagus2002exchangelike,vazquez2007energy,besse2021beyond}. However, we demonstrate for the first time that such a dipole can exhibit a dynamic response to external fields. In a GBR heterostructure, the interfacial quantum dipoles are found to relax to a pair of metastable states under external fields. As a result the system behaves like a ferroelectric that is switchable under gating and acquires nonlinear charge dynamics. The discovery of a new ferroelectric mechanism could shed light on the understanding of other unconventional hysteretic behaviors in van der Waals heterostructures, and is promising for the development of novel nanoelectronic memory devices.

\section*{Charge transfer heterostructure and hysteretic transport}

\begin{figure*}
 \includegraphics[width=1.8\columnwidth]{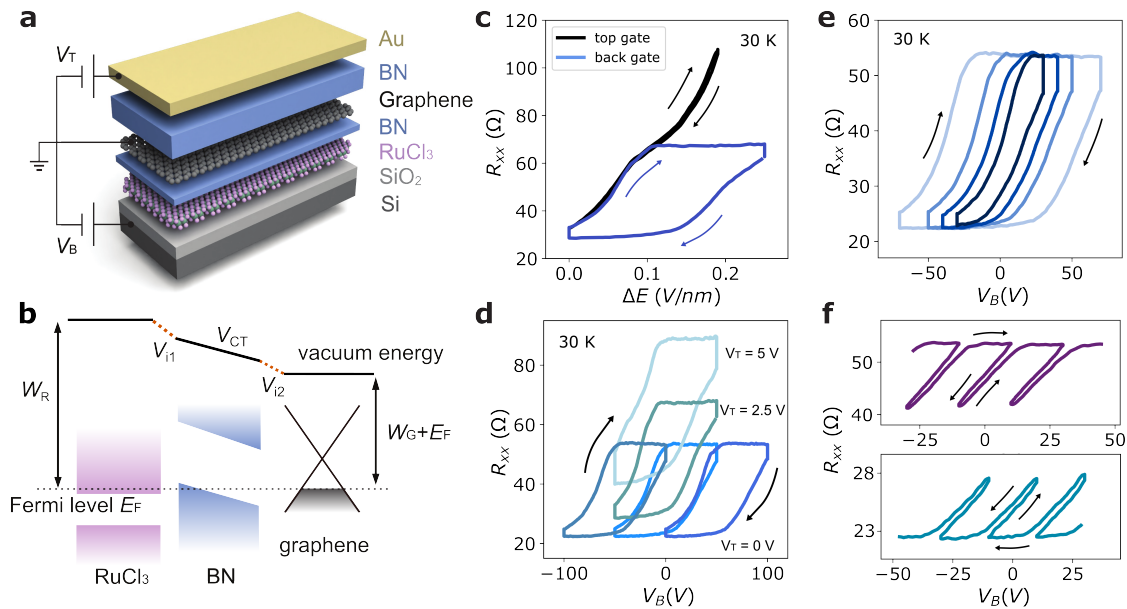}
 \caption{{\bf a,} Heterostructure of the charge transfer device with metal top and silicon bottom gates. {\bf b,} The band alignment of the graphene/$h$BN/$\alpha$-$\mathrm{RuCl_3}$ heterostructure. A difference in the work function ($W_R$ of $\alpha$-$\mathrm{RuCl_3}$ and $W_G$ of graphene) causes charge transfer doping in graphene (with Fermi level $E_F$) and band bending in $h$BN (with voltage drop $V_{CT}$). Interface polarization causes additional potential drops $V_{i1}$ and $V_{i2}$. {\bf c,} Top gate and bottom gate dependence of the graphene longitudinal resistance of device D1 (with a 3.5~nm $h$BN spacer and graphene $p$-doped to $5.1\times10^{12}$ cm$^{-2}$ under room temperature). {\bf d,} Tunability of the hysteresis loop position under top gate $V_T$ and bottom gate $V_B$. {\bf e,} Width tunability of the hysteresis loop set by the history of $V_B$ sweep. {\bf f,} Ratchet effect of the hysteresis near upper and lower resistance bounds. The saturated resistance can be immediately modulated wherever the bottom gate sweep direction is reversed.}
 \label{fig1}
\end{figure*}

Figure~\ref{fig1}a shows a cartoon schematic of a typical device. The heterostructure is assembled from exfoliated layers using the dry transfer technique and then etched into a dual-gated Hall bar geometry (Methods). The main feature of the GBR heterostructure is the top-down sequence consisting of monolayer graphene/multilayer $h$BN/multilayer $\alpha$-$\mathrm{RuCl_3}$. Voltage applied to the top gate, $V_{T}$, couples directly to the graphene layer through a multilayer $h$BN dielectric. The bottom gate electrode, $V_{B}$, consists of doped silicon separated from the $\alpha$-$\mathrm{RuCl_3}$ by a 300~nm native oxide layer.\\

In the equilibrium state, the large mismatch between the work functions of graphene ($W_G$) and $\alpha$-$\mathrm{RuCl_3}$ ($W_R$) gives rise to charge transfer and band bending across the heterostructure~\cite{mashhadi2019spin,zhou2019evidence,wang2020modulation,rizzo2022nanometer,balgley2022ultrasharp,rossi2023direct}, as illustrated in Fig.~\ref{fig1}b. As a result, graphene is $p$-doped with a Fermi level $E_F$ lying slightly below the Dirac point, and $\alpha$-$\mathrm{RuCl_3}$ is $n$-doped with a Fermi level pinned to the bottom of the conduction band, where the hybridized band structure possesses a large density of states (Methods and Supplementary Sec. II. A)~\cite{mashhadi2019spin}. The thickness of the $h$BN spacer between the graphene and $\alpha$-$\mathrm{RuCl_3}$ layers determines the magnitude of charge transfer~\cite{wang2020modulation}.\\

We first consider the longitudinal resistance measured in a GBR device with 3.5~nm thick $h$BN spacer, yielding an equilibrium density of $5.1\times10^{12}$ cm$^{-2}$. Figure~\ref{fig1}c highlights our main result. The device shows a typical response when varying the top gate bias (black curve), with the forward and reverse bias sweeps showing nearly identical behavior. In contrast, varying the bias on the bottom gate exhibits a strong hysteretic response. Interplay between the top and bottom gate potentials when varied simultaneously is shown in Fig.~\ref{fig1}d. The top gate bias shifts the average resistance value of the loop (vertical shift in the plot), but leaves the difference between the upper and lower resistance bounds unchanged. Together with Fig.~\ref{fig1}c, this suggests that the top gate can freely modulate the graphene resistance, whereas the bottom gate is only capable of modulating the channel resistance within a fixed range relative to the top gate induced average. The voltage boundaries of the hysteresis loop are independent of the top gate bias, and are instead determined entirely by the history of the bottom gate. Hysteresis appears only after the bottom-gated resistance reaches its upper or lower bounds. Furthermore, the hysteresis loop boundaries are set by wherever the bottom gate sweep direction is reversed, and do not depend on absolute voltages applied, or the specific carrier density in the graphene channel. This allows the loop position (horizontal shift in Fig.~\ref{fig1}d) and width (Fig.~\ref{fig1}e) to be broadly and dynamically tuned. Leveraging this tunability, an electronic ratchet effect~\cite{zheng2023electronic} can be demonstrated by sequentially sweeping the bottom gate, but restricting its range in each sweep between the upper and low resistance bounds, as shown in Fig.~\ref{fig1}f (and Supplementary Sec. III. B).

\section*{Interfacial quantum dipole}
\begin{figure*}
 \includegraphics[width=1.8\columnwidth]{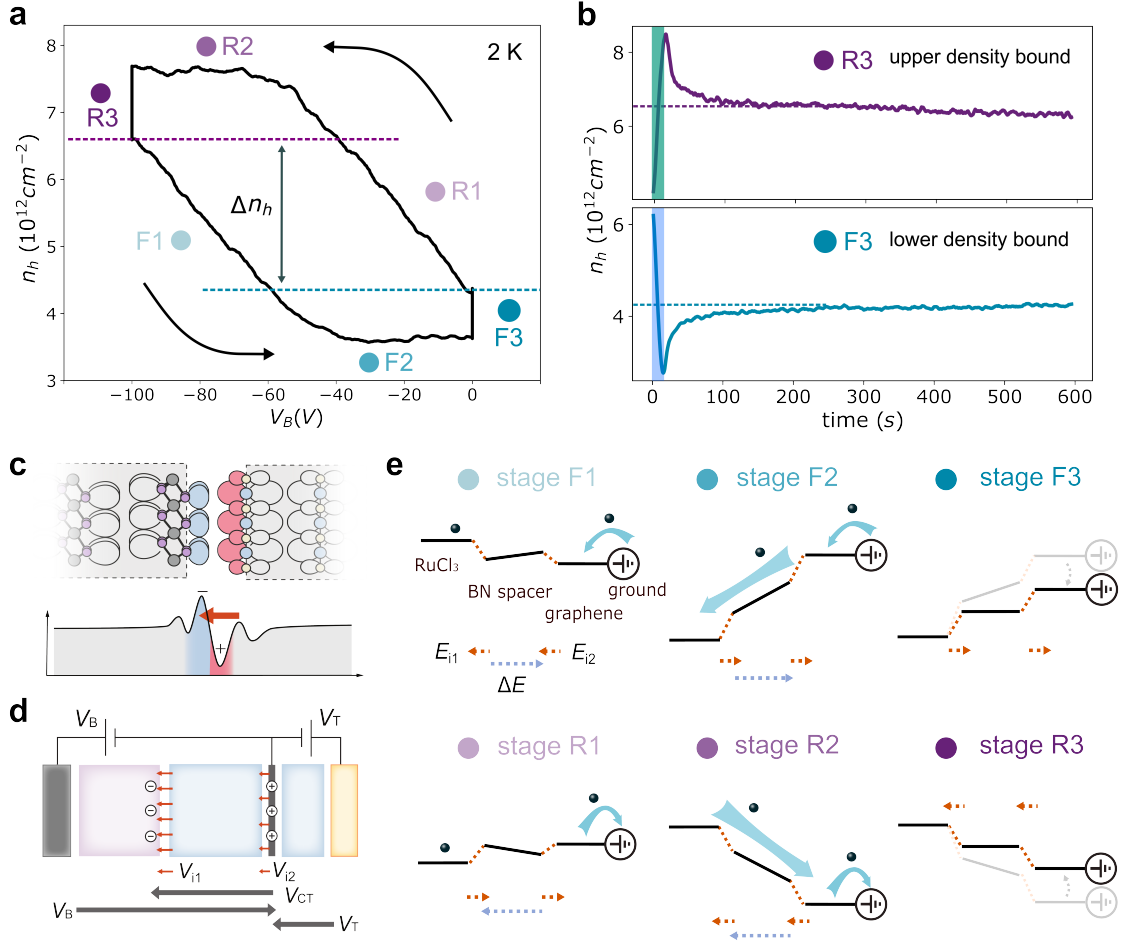}
 \caption{{\bf a,} The graphene Hall density $n_h$ (measured at 0.2 T) as a function of the bottom gate $V_B$ displays a hysteresis loop for device D2 (with a 4.5~nm $h$BN spacer and graphene $p$-doped to $5.4\times10^{12}$ cm$^{-2}$ under room temperature). Color dots mark different stages in a forward or reverse gate sweep. The purple (dark cyan) dashed line marks the upper (lower) density bound that the system relaxes to after each gate sweep. {\bf b,} Time dependence of the Hall density in a forward (lower panel) and reverse (upper panel) gate sweep. After each gate sweep (lasting 15~s marked by shaded regions), the Hall density is found to relax to and later stabilize at the upper or lower density bound indicated by dashed lines. {\bf c,} Illustration of the interfacial quantum dipole between $\alpha$-$\mathrm{RuCl_3}$ (left side) and $h$BN (right side) without charge transfer. Upper panel sketches the atomic orbitals at the interface and lower panel plots the corresponding electron density of states. Quantum exchange interactions deform local electronic orbitals which create an effective dipole (red arrow) across the interface. {\bf d,} Electric fields in the graphene/$h$BN/$\alpha$-$\mathrm{RuCl_3}$
 heterostructure associated with the interfacial potential drops $V_{i1}$ and $V_{i2}$, the charge transfer potential $V_{CT}$, the bottom gate $V_B$ and the top gate $V_T$. {\bf e,} Schematics of the electron potential, relative to the charge transfer equilibrium, for the six stages indicated in {\bf a}. The external electric field modulates the field strength across the heterostructure by $\Delta E$, which can switch the interfacial quantum dipoles and flip the directions of interfacial dipolar fields $E_{i1}$ and $E_{i2}$. The flow of electrons indicated by light cyan arrows only occurs between graphene and ground in stages F1 and R1 due to the potential barriers of the interfacial quantum dipoles, but can extend across the $h$BN spacer in later stages. In stages F3 and R3, the relative electron potential schematics in the saturated stages (light gray lines) relax back to the upper or lower density bound (solid black lines).}
 \label{fig2}
\end{figure*}

The observed hysteresis suggests that while the top gate couples to the graphene channel through the usual capacitor model, the bottom gate has a more complicated influence. To better understand the origin of the hysteresis loop we plot in Fig.~\ref{fig2}a the Hall density $n_h$ measured in the graphene layer at T = 2~K and B = 0.2~T while varying the bottom gate $V_B$. The response in Fig.~\ref{fig2} was measured from a second device with 4.5~nm thick $h$BN spacer and an equilibrium density of $5.4\times10^{12}$ cm$^{-2}$. In this measurement we sweep the gate at a rate of 0.8~V/s, first from from -100~V to 0~V (forward sweep) and then from 0~V to -100~V (reverse sweep). At the end of each sweep direction we hold the bias voltage for 5 minutes. The forward and reverse gate sweep each features three distinct stages. In stage F1, an increase of $V_B$ causes a linear decrease of the graphene hole density. The linear gating behavior indicates that in this stage the bottom gate couples directly to the graphene channel, unaffected by the $\alpha$-$\mathrm{RuCl_3}$ layer. As $V_B$ continues to increase, $n_h$ saturates to a nearly constant plateau (stage F2), where the graphene layer density shows nearly zero variation upon further gating. At the end of the sweep we note that the density slowly relaxes to a stable point (stage F3), indicated by the dark cyan dashed line, which we refer to as the lower density bound. Reversing the sweep direction shows opposite behavior -- $n_h$ increases linearly with decreasing $V_B$ in stage R1, followed by a saturation in stage R2 and a slow relaxation to a lower density than the stable value in stage R3, which we refer to as the upper density bound (purple dashed line). The density bounds represent a pair of metastable states that are independent of gate sweep parameters. We note that upon sweeping the gate more slowly, the plateau density approaches the long time equilibrium value without overshooting (Supplementary Sec. III. C). Fig.~\ref{fig2}b shows the time evolution of the graphene Hall density during stages F3 (lower panel) and R3 (upper panel) following rapid gate sweeps (6.6~V/s indicated by the shaded regions). The density consistently relaxes back to the metastable states marked by the dashed lines.\\

Through first principles calculations, we identify that the emergence of interfacial quantum dipoles play a crucial role in the dynamics of the gate response. Fig.~\ref{fig2}c upper panel shows a cartoon sketch of the atomic orbitals at a $h$BN/$\alpha$-$\mathrm{RuCl_3}$ interface. The lower panel plots the corresponding electron density based on density functional theory. Quantum exchange interactions distort the electron wavefunction and lead to an effective interfacial dipole (Methods and Supplementary Sec. II). This closely resembles the "pillow effect" - a theoretical model that has been used to explain the anomalous work function shift found at metal/organic interfaces ~\cite{bagus2002exchangelike,vazquez2007energy} where in a similar way, an interfacial dipole results from distortion of the electron wavefunctions rather than charge transfer. Key to establishing the analogous pillow effect in our device is a clean van der Waals interface devoid of charge traps or chemical reactivity~\cite{wang2013one}. Previous studies of the pillow effect considered only its static influence~\cite{besse2021beyond,rizzo2023polaritonic,rossi2023direct}. Here we demonstrate for the first time that applying an external electric field continuously tunes the interfacial quantum dipole.\\

Figure~\ref{fig2}d shows the individual dipole potentials that contribute to the net electric field profile across the GBR structure under gating. While the top gate $V_T$ couples to the graphene channel directly, the electric field induced by the bottom gate $V_B$ penetrates across the GBR heterostructure, enabling modulation of potential drops both in the spacer $V_{CT}$ and at interfaces $V_{i1}, V_{i2}$. Fig.~\ref{fig2}e shows cartoon schematics that illustrate how the calculated electrostatic potential configuration varies for each stage of the hysteresis loop (relative to the charge transfer equilibrium at room temperature, see Supplementary Sec. II). In the presence of interfacial dipole potentials (orange dashed lines), the charge flowing between graphene and $\alpha$-$\mathrm{RuCl_3}$ can either be impeded or assisted. In going from stage F1 to F2, there are initial potential barriers set up across the interfaces, preventing electrons to flow from graphene to $\alpha$-$\mathrm{RuCl_3}$. The effect of the bias is therefore to increase the electric field $\Delta E$ across the $h$BN spacer and to raise the graphene electron potential (leading to graphene hole discharging), until a point where the interfacial potential drops are reversed and electrons are allowed to flow from graphene to $\alpha$-$\mathrm{RuCl_3}$ (indicated by the light cyan arrow). Similarly, in going from stage R1 to R2, there is an initial hole barrier across the heterostructure, leading to graphene hole charging with decreased external bias until the interfacial potential drops flip back, and charges can flow in the opposite direction.\\ 

At the end of each gate sweep, a relaxation process appears in which charges redistribute until the internal field across the $h$BN spacer $\Delta E$ flattens out relative to charge transfer equilibrium at room temperature. The system then settles into a metastable configuration with the interfacial quantum dipole potential trapped at finite energies (stages F3 or R3), defining the upper or lower density bound.\\

The metastable state satisfies the relationship $W_G - W_R = eV_{CT} + eV_{i1} + eV_{i2} + E_F(n_h)$, where $e$ is the (positive) elementary charge and $E_F$ is the graphene Fermi energy as a function of density $n_h$. The observed difference between the metastable density bounds, $\Delta n_h = 2.2\times10^{12} cm^{-2}$ (marked in Fig.~\ref{fig2}a), therefore suggests a variation in interfacial quantum dipole potentials $\Delta V_{i1} + \Delta V_{i2} = \pm 0.26$~eV, in close agreement with the value of $\pm 0.3$~eV obtained from first principles calculations (Supplementary Sec. II. E).

\section*{Hysteresis dynamics}

\begin{figure*}
 \includegraphics[width=1.7\columnwidth]{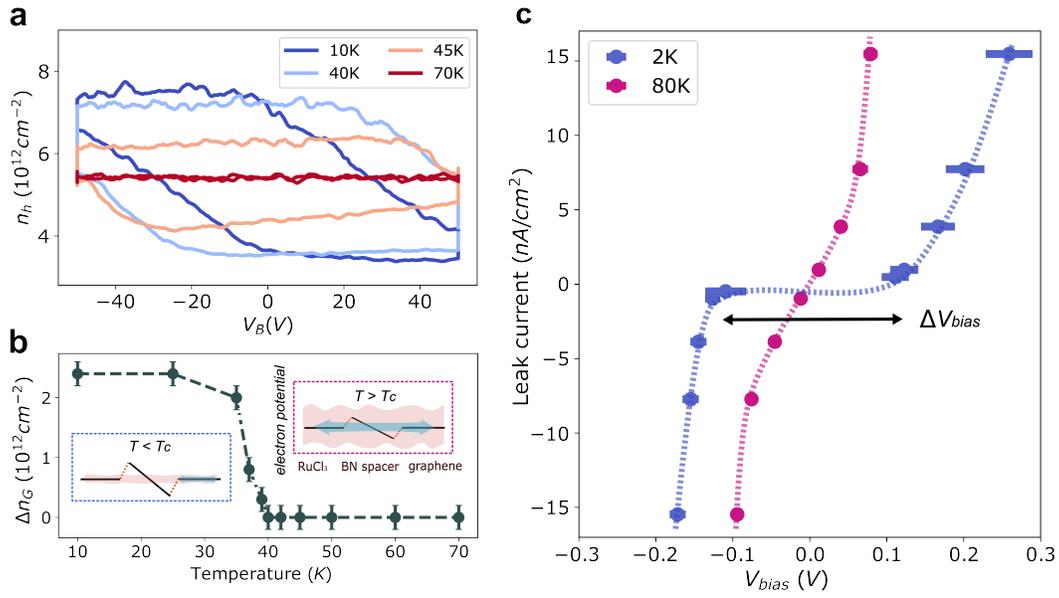}
 \caption{{\bf a,} Temperature dependence of the hysteresis loop of the graphene Hall density of device D2. {\bf b,} Temperature dependence of the difference between the metastable density bounds, $\Delta n_{h}$. Insets: schematics of the electron potential at different temperatures. Below transition temperature the interfacial potentials prevent charge transfer across the heterostructure, while at higher temperatures thermal fluctuations assist the current flow independent of the interfacial quantum dipole orientations. {\bf c,} The extracted nonlinear I-V characteristics of the charging dynamics across graphene/$h$BN/$\alpha$-$\mathrm{RuCl_3}$ for device D1.}
 \label{fig3}
\end{figure*}

Figure~\ref{fig3}a shows the temperature dependence of the hysteresis loop for the same device measured in Fig.~\ref{fig2}. As shown in Fig.~\ref{fig3}b, the difference between the metastable density bounds $\Delta n_h$ is mostly temperature independent in the low regime, and then sharply collapses around T = 40~K, beyond which the bottom gate becomes fully screened. The sharply defined critical temperature that separates the hysteretic and fully screened behaviors suggests that the associated interfacial dipole potential barriers regulate tunneling current flow at low temperatures but can be overcome by thermal fluctuations at higher temperatures, as illustrated in the insets of Fig.~\ref{fig3}b.\\

The nonlinear dynamics embedded in the hysteretic transport can be extracted from the gate sweep speed dependence. The saturated stages in the hysteresis loop signifies a balance between the total doping rate of the system (set by the bottom gate sweep rate) and the charge tunneling rate across the GBR heterostructure. By tracking the density values of the saturated stages under different gate sweep rates, we can obtain the I-V characteristic of the charge tunneling across the GBR heterostructure (Supplementary Sec. III. C). Figure~\ref{fig3}c presents the I-V characteristics at T = 2~K and 80~K for the device measured in Fig.~\ref{fig1}. The charge dynamics is clearly nonlinear with a gap-like feature $\Delta V_{bias}$ appearing at 2~K but closing at 80~K, consistent with the temperature dependence of the hysteresis loop. The size of the ‘gap’ is directly related to the splitting between the two density bounds $\Delta n_{h}$. At 2~K, this ‘gap’ is the result of the potential barrier set up by the interfacial quantum dipoles, which limits charge tunneling across the GBR heterostructure. First principles calculations give an estimate of the barrier energy associated with each interfacial quantum dipole to be around 0.1 - 0.2~eV (Supplementary Sec. II. E), in good agreement to the ‘gap’ size $\Delta V_{bias} = 0.22$~eV in Fig.~\ref{fig3}c. At higher temperatures, thermal energy strongly enhances the tunneling through the potential barrier (Fig.~\ref{fig3}b insets), effectively shorting $\alpha$-$\mathrm{RuCl_3}$ to electrical contacts through graphene and resulting in complete screening of the bottom gate (Fig.~\ref{fig3}a).

\section*{Discussions}

The model of dynamic interfacial quantum dipoles successfully accounts for the presence of tunable hysteretic transport, its temperature dependence, and the nonlinear charge dynamics. Alternative mechanisms for hysteresis are excluded as outlined below: 1) Charge traps: mobile ions and defects in the dielectric can contribute to hysteretic behaviors~\cite{wang2010hysteresis,joshi2010intrinsic,liao2010hysteresis}, and $\alpha$-$\mathrm{RuCl_3}$ may have in-gap localized states that serve as charge traps~\cite{rossi2023direct}. However, hysteresis caused by thermally activated charge traps is typically suppressed under low temperatures~\cite{kaushik2017reversible}, which contradicts our observation. In addition, the wide-range voltage tunability and stability of the observed hysteresis loop imply the presence of a large charge reservoir (Supplementary Sec. III. A), which are not accounted for by the disordered nature of charge traps, particularly under heavily doped conditions. 2) Nonlinear tunneling across $h$BN barrier: hysteresis is widely reported in NRAM devices with $h$BN tunneling barriers~\cite{liu2021ultrafast,wu2021atomically}. However, our $h$BN spacer is much thinner than the typical $h$BN dielectric used for floating gate memory devices~\cite{liu2021ultrafast,wu2021atomically}. Under the charge transfer potential, the thin $h$BN spacer becomes effectively transparent compared to the observed slow charge dynamics~\cite{fowler1928electron,frenkel1938pre}, resulting in substantial leakage currents that prevent the realization of NRAM functionality. Furthermore, the observed hysteresis exhibits no dependence on $h$BN thickness and persists even without $h$BN spacer. 3) Phase transitions of $\alpha$-$\mathrm{RuCl_3}$: interactions may open a correlated gap in the conduction band of $\alpha$-$\mathrm{RuCl_3}$ under low temperatures. We argue that phase transitions in $\alpha$-$\mathrm{RuCl_3}$ are not likely to play a role in the temperature dependence of the hysteresis and the corresponding I-V characteristics. The onset temperature of hysteresis varies from T = 40 to 80~K in our devices (Supplementary Sec. III. C), making it unlikely to be associated with the reported magnetic transitions in bulk $\alpha$-$\mathrm{RuCl_3}$ which are typically observed at temperatures around T = 10~K and 100~K~\cite{banerjee2016proximate,do2017majorana}.\\

Interfacial quantum dipoles are expected to be ubiquitous across van der Waals interfaces~\cite{rizzo2023polaritonic,rossi2023direct,besse2021beyond}, but typically yield negligible effects in transport. Indeed, we do not observe similar hysteresis in our devices in the absence of $\alpha$-$\mathrm{RuCl_3}$, pointing to its critical role. We conjecture that the presence of a narrow electronic band in $\alpha$-$\mathrm{RuCl_3}$ is the key feature. By pinning the Fermi energy while allowing the penetration of external fields, $\alpha$-$\mathrm{RuCl_3}$ serves as a pivot for modulating the potential profile across the GBR heterostructure.\\

The GBR heterostructure that we study here resembles a floating gate transistor in which the bottom gate serves as the control gate and $\alpha$-$\mathrm{RuCl_3}$ acts as a charge trapping layer~\cite{zhang2023van}, but with the distinctive feature of a tunable barrier controlled by the dynamic history of the interfacial quantum dipoles. We anticipate that similar charge transfer heterostructures featuring semiconducting monolayers in place of graphene will exhibit similar behavior, providing new opportunities for dynamically tunable semiconductor technologies such as programmable memory cells, tunnel field-effect transistors and customizable neural networks~\cite{yin2021emerging}.

%%%%%%%%%%%%%%%%%%%%%%%%%%%%%%%%%%%%%%%%%%%%%%%%%%%%%%%
%%%%%%%%%%%%%%%%%%%%%%%%%%%%%%%%%%%%%%%%%%%%%%%%%%%%%%%
%%%%%%%%%%%%%%%%%%%%%%%%%%%%%%%%%%%%%%%%%%%%%%%%%%%%%%%
\clearpage

\section*{Methods}
\subsection*{Device fabrication}
Flakes of graphene, graphite, $h$BN, and $\alpha$-$\mathrm{RuCl_3}$ are mechanically exfoliated onto Si/$\mathrm{SiO_2}$ substrates (with 285 nm or 90 nm oxide thickness) and identified by optical contrast. To reduce the adhesion between $\alpha$-$\mathrm{RuCl_3}$ and the substrate, Si/$\mathrm{SiO_2}$ substrates are precoated with 1-dodecanol for 5 min at 160 $\mathrm{^oC}$ and rinsed with isopropyl alcohol. The GBR heterostructure with $h$BN dielectrics are assembled using the van der Waals dry transfer technique. The thin $h$BN spacer enables graphene to be remotely and controllably doped by $\alpha$-$\mathrm{RuCl_3}$ while maintaining its high quality. The flakes are picked up in sequence by polycarbonate (PC) transfer polymer, rested on a poly dimethyl siloxane (PDMS) stamp. The final stack is released from the polymer stamp onto Si/$\mathrm{SiO_2}$ substrate at 180 $\mathrm{^oC}$.

Because $\alpha$-$\mathrm{RuCl_3}$ is sensitive to chemicals including acetone and electron beam resists, the device fabrication protocol is carefully designed such that $\alpha$-$\mathrm{RuCl_3}$ remains encapsulated throughout the process. In particular, half-edge Ti/Pd/Au contacts are made to graphene without etching the thin $h$BN spacer. More details are provided in Supplementary Information. In the first step, windows for contacting graphene are opened using electron beam lithography and reactive ion plasma etching. A sequence of $\mathrm{O_2}$, $\mathrm{SF_6}$ and short $\mathrm{O_2}$ plasma is used to clean the polymethyl methacrylate (PMMA) residuals and etch the stack down to the intermediate $h$BN spacer. Second, electron beam lithography and electron beam evaporation are used to deposit Ti/Pd/Au leads which contact graphene through the windows opened in the first step. The final etching step defines the Hall bar geometry. The electron beam resist with the predefined pattern is left on the top surface of the device so $\alpha$-$\mathrm{RuCl_3}$ is not exposed to any solvent. 

\subsection*{Transport measurements}
We measured electronic transport properties of graphene in a variable-temperature dry fridge with a base temperature of 1.5 K. Four- and two-terminal resistance measurements were carried out using a low-frequency lock-in technique at frequencies ranging from 17 Hz to 37 Hz. Graphene is grounded and current biased ranging from 10 nA to 100 nA. $\alpha$-$\mathrm{RuCl_3}$ remains floated and does not directly connect to any contact. Metal top gates and silicon bottom gates are d.c. biased. Seven devices were measured all showing similar hysteresis at low temperatures (Supplementary Sec. III. D).

\subsection*{Electrostatic model}
To describe the equilibrium configuration and charge dynamics of the GBR heterostructure, we consider an electrostatic model accounting for the bottom gate, the charge transfer dipole, and the interface polarizations (see Extended Data Fig.~\ref{fig:electrostatic_model}). For simplicity, we assume a one-dimensional model, which can be obtained by averaging the charge density and electric potential over the directions perpendicular to the stacking direction.

To describe the equilibrium charge configuration, we first assume that there is no polarization at the interfaces. Under these conditions, the work function difference $\Delta W$ between graphene and $\alpha$-$\mathrm{RuCl_3}$ has to be compensated by a charge transfer dipole. This dipole arises from electron transfer from graphene to $\alpha$-$\mathrm{RuCl_3}$, leading to a reduction of the Fermi level and hole doping of graphene. Denoting the electric potential corresponding to this dipole by $V_{\rm CT}$, we have the balance condition $\Delta W + eV_{\rm CT} + E_F = 0$, where the work functions of $\alpha$-$\mathrm{RuCl_3}$ and graphene are measured from the bottom of the conduction band and the Dirac cone, respectively, and $E_F$ is the Fermi level of graphene measured down from the Dirac cone. The charge transfer dipole sets up an internal electric field $E_{\rm int} = - \partial_x V_{\rm CT}$ across the $h$BN structure, and leads to a flat potential energy between the $\alpha$-$\mathrm{RuCl_3}$ and graphene Fermi levels, the former of which is pinned to the conduction band minimum.

We now consider the effects of finite interface polarizations. From first principles calculations (see Supplementary Sec. II), we find finite potential drops $V_{i1}$ and $V_{i2}$ across the $\alpha$-$\mathrm{RuCl_3}$-$h$BN and $h$BN-graphene interfaces, respectively. The electrostatic balance equation then becomes
\begin{align}
 \Delta W + eV_{\rm CT} + E_F + eV_{i1} + eV_{i2} = 0,
\end{align}
where $V_{i1}$ and $V_{i2}$ are assumed to be independent of the charge transfer density. This assumption has been verified via first principles calculations, and holds to a very good degree (see Supplementary Fig. S4). In contrast, the interfacial potentials are found to depend on the externally applied electric field $E_{\rm ext}$ (see Supplementary Fig. S6). Given the quantities $\Delta W$, $V_{i1}$ and $V_{i2}$, the equation above determines the charge transfer density and the values of $V_{\rm CT}$ and $E_F$.

With the interface polarizations included, the potential energy landscape has the zig-zag form as shown in Extended Fig.~\ref{fig:electrostatic_model}, with the potentials on the left and right hand sides are assumed to be at the $\alpha$-$\mathrm{RuCl_3}$ and graphene Fermi levels, respectively. In contrast to the earlier situation, there is now a finite potential energy difference across the $h$BN spacer, coming from the difference between the work function mismatch and the charge transfer dipole, as well as the two regions of increasing potential at the interfaces.

From first principles calculations (see below and Supplementary Sec. II), we find a total potential drop across the heterostructure of $eV_{\rm tot} = e(V_{\rm CT} + V_{i1} + V_{i2}) = \Delta W - E_F = 1.61$ eV, together with the interfacial dipole potentials $eV_{i1} = 0.74$ eV and $eV_{i2} = 0.14$ eV. Using $E_F = \hbar v_F \sqrt{\pi n_h} = 0.31$ eV, relative to the Dirac cone, as obtained with $v_F = 1.15 \times 10^8$ cm s$^{-1}$ and $n_h = 5.5 \times 10^{12}$ cm$^{-2}$, the work function difference is $\Delta W = 1.92$ eV and the charge transfer potential is $eV_{\rm CT} = 0.72$ eV. This is close to the value $eV_{\rm CT} = e^2\sigma d/\epsilon_r\epsilon_0 = 0.85$ eV expected from treating the heterostructure as a parallel plate capacitor, and using $\sigma = 5.1 \times 10^{12}$ cm$^{-2}$, $\epsilon_r = 3.76$~\cite{Bokdam2011,Laturia2018} and d = 3.5 nm. We note that the electric field across the $h$BN layers is $E_{\rm int} = \partial_d V_{\rm CT} =  e\sigma/\epsilon_r\epsilon_0 = 0.21$ V/nm.

When an external electric field $E_{\rm ext}$ is applied across the heterostructure, it has two effects: The first is to modify the electric field across the $h$BN structure, such that the total electric field is now $E_{\rm tot} = E_{\rm int} + E_{\rm ext}$. The second effect is to change the values of the interface polarizations, as a result of charge reconfigurations within each sub-system (see Supplementary Sec. II). Assuming a heterostructure of total height $h = 300$ nm (where the dominant part consists of the SiO$_2$ substrate), we note that an applied bias of $V_{\rm ext} = 25$ V (as used in our measurements) corresponds to an electric field $E_{\rm ext}$ = 0.08 V/nm. We fix the zero of potential to the $\alpha$-$\mathrm{RuCl_3}$ Fermi level, which to a very good approximation is pinned to the conduction band minimum due to the large $\alpha$-$\mathrm{RuCl_3}$ density of states (see Supplementary Sec. II).

\subsection*{Temperature dependence and density swing}
The electrostatic model is now used to estimate the temperature dependence and density swing (the difference between upper and lower density bounds) of the hysteresis loop. To obtain the critical temperature below which the system becomes hysteretic, we note that the hysteresis arises from a competition between charge doping and charge leakage across the heterostructure. Starting from the relaxed metastable state of Fig.~\ref{fig2}b, thermal fluctuations allow charges to move across the heterostructure until an electric field barrier of height $U$ is set up across the $h$BN spacer. At $T = 0$, any barrier $U > 0$ is sufficient to prevent further charge transfer, while at finite temperature a barrier $U \approx k_BT$ is needed (see Fig.~\ref{fig3}b insets). The lifetime of the metastable state follows from the Arrhenius formula $\tau = (\hbar/\epsilon) e^{U/k_BT}$~\cite{Hanggi1990}, where $\epsilon$ is a characteristic energy scale. Here we use $\epsilon = 0.1$ eV, but the results are insensitive to this number since $\tau$ is dominated by the strong exponential behavior. As the maximal value of $U$ is set by the interfacial dipoles which gives $U_{\rm max} \approx 0.1 - 0.2$ eV (see Supplementary Sec. II. E), we find a critical temperature $T_c \approx 50$ K above which $\tau$ becomes comparable with the charge doping lifetime ($\tau_d \sim 10$ s). Since $U$ is effectively a function of temperature, we expect the transition from the high temperature non-hysteretic regime ($T > T_c$), to the low temperature hysteretic regime ($T < T_c$) to be sharp.

To estimate the density swing, we calculate the difference in hole density between the two metastable states. We denote the Fermi energy of the high temperature regime by $\epsilon_F$, and let $V_d$ be the total potential drop across the heterostructure at $T = 0$ (see Figs.~\ref{fig2}b and \ref{fig2}e). We can then write the density swing as $\Delta n = n(\epsilon_F) eV_d/\epsilon_F - n(\epsilon_F)(-eV_d)/\epsilon_F = n(\epsilon_F) (2V_d/\epsilon_F)$, where we have used that the graphene density is a linear function of the external bias~\cite{Kim2012}. Using $\epsilon_F = 0.31$ eV, corresponding to a charge density $n(\epsilon_F) = 5.5 \times 10^{12}$ cm$^{-2}$, and a potential drop of $V_d = 0.1$ eV, gives a density swing of $\Delta n = 3.5 \times 10^{12}$ cm$^{-2}$. This is in qualitative agreement with our measurements, and is likely an upper bound since we have neglected additional processes (such as charge tunneling) that could lead to additional charge leakage in the low temperature limit.

\subsection*{First principles parameterization}
To evaluate the electric field dependence of the charge transfer and interfacial dipoles, we performed density functional theory (DFT) calculations. The simulations were done by first calculating the charge density and electric potential of each subsystems separately (i.e., $\alpha$-$\mathrm{RuCl_3}$, $h$BN and graphene), and then subtracting these quantities from those of the full heterostructure to obtain the density and potential differences~\cite{Bokdam2011}
\begin{align}
 \Delta n &= n_{{\rm RuCl}_3-h{\rm BN}-{\rm Gr}} - n_{{\rm RuCl}_3} - n_{h{\rm BN}} - n_{\rm Gr} \\
 \Delta V &= V_{{\rm RuCl}_3-h{\rm BN}-{\rm Gr}} - V_{{\rm RuCl}_3} - V_{h{\rm BN}} - V_{\rm Gr}. \nonumber
\end{align}
In all calculations, we used a heterostructure with three $h$BN layers between single layers of $\alpha$-$\mathrm{RuCl_3}$ and graphene. To determine the values of the interfacial dipoles, we also performed analogous calculations for the separate $\alpha$-$\mathrm{RuCl_3}$-$h$BN and $h$BN-Gr interfaces. Calculations were performed both in the zero-field state ($E_{\rm ext} = 0$), and at finite electric fields.

The DFT calculations were performed with the {\sc Octopus} electronic structure code~\cite{TancogneDejean2017,TancogneDejean2020}, within the local density approximation, using norm conserving pseudopotentials from the Pseudodojo project. To treat the local correlations on the Ru atoms, we used the hybrid DFT+$U$ functional ACBN0~\cite{TancogneDejean2017}, with a value of $U = 2$ eV. The calculations were performed in a $2 \times 2$ hexagonal $\alpha$-$\mathrm{RuCl_3}$ supercell, with a length of $12.3$ {\AA} along each lattice vector, corresponding to a $5 \times 5$ supercell for $h$BN and graphene. We used mixed boundary conditions, periodic in the in-plane directions and open in the out-of-plane direction, with a vacuum region of $35$ {\AA } to ensure convergence in the out-of-plane direction. A $3\times 3$ $k$-point grid and a real-space grid spacing of $0.3$ Bohr were employed. The electric field was included via the modern theory of polarization, using a single-point calculation of the Berry phase. To obtain the relaxed metastable configurations following a sudden switch-off of the external electric field, we performed imaginary time propagation with a time step of $0.01$ a.u., starting from the density $n(E_{\rm ext})$ at finite $E_{\rm ext}$.

\bibliography{references}
% \putbib[references]
% \end{bibunit}

%%%%%%%%%%%%%%%%%%%%%%%%%%%%%%%%%%%%%%%%%%%% Extended figures

\setcounter{figure}{0}
\renewcommand{\figurename}{Extended Fig.}
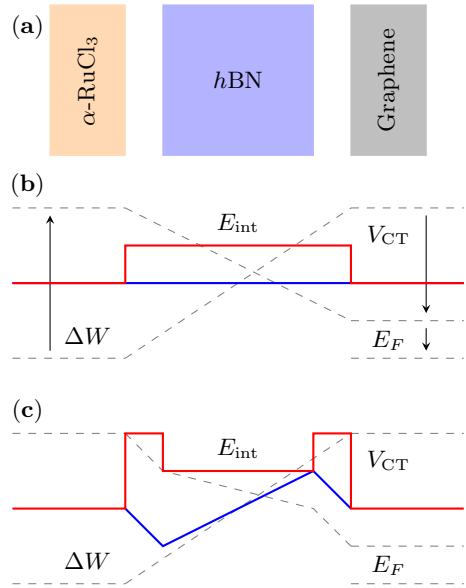
\begin{figure*}
 \centering
 \begin{tikzpicture}

  % Materials
  \draw[orange!30!white,fill=orange!30!white] (-2.0,0.2) rectangle (-1.0,2.2);
  \draw[blue!30!white,fill=blue!30!white] (-0.5,0.2) rectangle ( 1.5,2.2);
  \draw[lightgray,fill=lightgray] ( 2.0,0.2) rectangle ( 3.0,2.2);

  % Labels
  \node[rotate=90] at (-1.5, 1.20) {$\alpha$-$\mathrm{RuCl_3}$};
  \node[rotate=0]  at ( 0.5, 1.2) {$h$BN};
  \node[rotate=90] at ( 2.5, 1.2) {Graphene};

  % Work function
  \draw[dashed,gray] (-2.5,-2.5) -- (-1.0,-2.5);
  \draw[dashed,gray] (2.0,-0.5) -- (3.5,-0.5);
  \draw[dashed,gray] (-1.0,-2.5) -- (2.0,-0.5);

  % Charge transfer dipole
  \draw[dashed,gray] (-2.5,-0.5) -- (-1.0,-0.5);
  \draw[dashed,gray] ( 2.0,-2.0) -- (3.5,-2.0);
  \draw[dashed,gray] (-1.0,-0.5) -- (2.0,-2.0);
  \draw[dashed,gray] ( 2.0,-2.5) -- (3.5,-2.5);

  \draw[->,>=stealth] (-2.0,-2.4) -- (-2.0,-0.6);
  \draw[<-,>=stealth] ( 3.0,-1.9) -- ( 3.0,-0.6);
  \draw[<-,>=stealth] ( 3.0,-2.4) -- ( 3.0,-2.1);

  % Potential and electric field
  \draw[thick,blue] (-2.5,-1.5) -- (3.5,-1.5);
  \draw[thick,red]  (-2.5,-1.5) -- (-1.0,-1.5) -- (-1.0,-1.0) -- (2.0,-1.0) -- (2.0,-1.5) -- (3.5,-1.5);

  % Labels
  \node at (-1.5,-2.20) {$\Delta W$};
  \node at ( 2.5,-0.86) {$V_{\rm CT}$};
  \node at ( 2.5,-2.26) {$E_F$};
  \node at ( 0.5,-0.7) {$E_{\rm int}$};

  % Work functions and charge transfer dipole
  \draw[dashed,gray] (-2.5,-5.5) -- (-1.0,-5.5);
  \draw[dashed,gray] ( 2.0,-3.5) -- ( 3.5,-3.5);
  \draw[dashed,gray] (-1.0,-5.5) -- ( 2.0,-3.5);

  % Work functions and charge transfer dipole
  \draw[dashed,gray] (-2.5,-3.5) -- (-1.0,-3.5);
  \draw[dashed,gray] ( 2.0,-5.0) -- (3.5,-5.0);
  \draw[dashed,gray] (-1.0,-3.5) -- (-0.5,-4.0) -- (1.5,-4.5) -- (2.0,-5.0);
  \draw[dashed,gray] ( 2.0,-5.5) -- (3.5,-5.5);

  % Potential and electric field
  \draw[thick,blue] (-1.0,-4.5) -- (-0.5,-5.0) -- ( 1.5,-4.0) -- ( 2.0,-4.5);
  \draw[thick,red]  (-2.5,-4.5) -- (-1.0,-4.5) -- (-1.0,-3.5) -- (-0.5,-3.5) -- (-0.5,-4.0) -- (1.5,-4.0) -- (1.5,-3.5) -- (2.0,-3.5) -- (2.0,-4.5) -- (3.5,-4.5);

  \node at (-1.5,-5.20) {$\Delta W$};
  \node at ( 2.5,-3.86) {$V_{\rm CT}$};
  \node at ( 2.5,-5.26) {$E_F$};
  \node at ( 0.5,-3.7) {$E_{\rm int}$};

  \node at (-2.3, 1.9) {({\bf a})};
  \node at (-2.3,-0.2) {({\bf b})};
  \node at (-2.3,-3.2) {({\bf c})};

 \end{tikzpicture}
 \caption{Equilibrium electrostatic model. ({\bf a}) Schematic of the stacking of the materials into a charge transfer heterostructure. ({\bf b}) Equilibrium potential landscape without interface polarizations. The blue lines show the potential energy and the red line the electric field. ({\bf c}) Equilibrium potential landscape with interface polarizations. The blue lines show the potential energy and the red line the electric field.}
 \label{fig:electrostatic_model}
\end{figure*}

\section*{Acknowledgments}
This research is primarily supported by Programmable Quantum Materials, an Energy Frontier Research Center funded by the US Department of Energy, Office of Science, Basic Energy Sciences, under award no. DE-SC0019443. Device fabrication was supported in part by the Columbia University Materials Science and Engineering Research Center (MRSEC), through NSF grants DMR-1420634 and DMR-2011738. E.V.B. acknowledges funding from the European Union's Horizon Europe research and innovation program under the Marie Sk{\l}odowska-Curie grant agreement No 101106809. This work was supported in part by the European Research Council (ERC-2024-SyG-101167294; UnMySt), the Cluster of Excellence: Advanced Imaging of Matter (AIM), and Grupos Consolidados UPV/EHU (IT1249-19). We acknowledge support from the Max Planck-New York City Center for Non-Equilibrium Quantum Phenomena. The Flatiron Institute is a division of the Simons Foundation. D.G.M. and M.C. acknowledge support from the Gordon and Betty Moore Foundation’s EPiQS Initiative, Grant GBMF9069. K.W. and T.T. acknowledge support from the JSPS KAKENHI (Grant Numbers 21H05233 and 23H02052) , the CREST (JPMJCR24A5), JST and World Premier International Research Center Initiative (WPI), MEXT, Japan.

\section*{Author contributions}
Z.L. and D.S. fabricated the devices. Z.L. performed the
electronic transport measurements and analyzed the data. Z.L., J.P. and C.R.D. discussed the experimental results. C.R.D. supervised the project. E.V.B. and A.R. performed the theoretical work. K.W. and T.T. grew the hexagonal boron nitride crystals. M.C. grew the $\alpha$-$\mathrm{RuCl_3}$ crystals under the supervision of D.G.M. Z.L., E.V.B., A.R. and C.R.D. wrote the manuscript with input from all authors.

\clearpage
%\onecolumngrid

%%%%%%%%%%%%%%%%%%%%%%%%%%%%%%%%%%%%%%%%%%%%%%%%%%%%%%%
%%%%%%%%%%%%%%%%%%%%%%%%%%%%%%%%%%%%%%%%%%%%%%%%%%%%%%%
%%%%%%%%%%%%%%%%%%%%%%%%%%%%%%%%%%%%%%%%%%%%%%%%%%%%%%%

\end{document}